\documentclass{aa}  

\usepackage{amsmath}
\usepackage{amssymb}
\usepackage[]{graphicx}
\usepackage{subeqnarray}
\usepackage[varg]{txfonts}
\usepackage[version=4]{mhchem}
\usepackage{enumerate}
\usepackage{cases}
\usepackage{mathrsfs,amssymb}
\usepackage[usenames]{color}
\usepackage[breaklinks,pdfnewwindow=true,colorlinks=true,citecolor=blue]{hyperref}

\definecolor{orange}{rgb}{1,0.5,0}

\definecolor{pink}{rgb}{0.858, 0.188, 0.478}

\definecolor{darkgreen}{rgb}{0,0.5,0}

\definecolor{lightgreen}{rgb}{0,0.5,0}

\begin{document}

\title{Broadband spectroscopy of astrophysical ice analogues}
\subtitle{III. Scattering properties and porosity of
\ce{CO} and \ce{CO2} ices}

\author{
    A.A.~Gavdush \inst{\ref{GPI}},
    A.V.~Ivlev \inst{\ref{MPE}}\thanks{The two authors contributed equally to this work.},
    K.I.~Zaytsev \inst{\ref{GPI}},
    V.E.~Ulitko \inst{\ref{PM}},
    I.N.~Dolganova \inst{\ref{GPI}},
    S.V.~Garnov \inst{\ref{GPI}},
    B.M.~Giuliano\inst{\ref{MPE}},
    and P.~Caselli \inst{\ref{MPE}}
}

\institute{
    Prokhorov General Physics Institute of
    the Russian Academy of Sciences, 
    119991 Moscow, Russia
    \label{GPI}
    \and
    Max-Planck-Institut f\"ur Extraterrestrische Physik,
    Gießenbachstraße 1, Garching, 85748, Germany \\\email{ivlev@mpe.mpg.de}
    \label{MPE}
    \and
    Department of Engineering Physics,
    Polytechnique Montréal,
    Montreal, Quebec, H3C 3A7, Canada
    \label{PM}
}

\date{Received - , 2025;
Accepted - , 2025}

\authorrunning{A.A.~Gavdush~et~al.}

\titlerunning{Scattering properties and porosity}

\date{Received 2025; accepted  2025}   

    
\abstract
    {The quantification of the terahertz (THz) and IR
    optical properties of astrophysical ice analogs, which have
    different molecular compositions, phases, and structural properties, is required to model both the continuum emission by the dust grains covered with thick icy mantles and the radiative transfer in the dense cold regions of the interstellar medium.}
    {We developed a model to define a relationship between the THz--IR response and the ice porosity. It includes the reduced effective optical properties of porous ices and the additional wave extinction due to scattering on pores. The model is applied to analyze the measured THz--IR response of \ce{CO} and \ce{CO2} laboratory ices and to estimate their scattering properties and porosity.}
    {Our model combines the Bruggeman effective medium theory, the Lorentz-Mie and Rayleigh scattering theories, and the radiative transfer theory to analyze the measured THz--IR optical properties of laboratory ices.}
    {We apply this model to show that the electromagnetic-wave scattering in studied laboratory ices occurs mainly in the Rayleigh regime at frequencies below $32$~THz. We conclude that pores of different shapes and dimensions can be approximated by spheres of effective radius. By comparing the measured broadband response of our laboratory ices with those of reportedly compact ices from earlier studies, we quantify the scattering properties of our \ce{CO} and \ce{CO2} ice samples. Their porosity is shown to be as high as $15$\% and $22$\%, respectively. Underestimating the ice porosity in the data analysis leads to a proportional relative underestimate of the THz--IR optical constants.}
    {The scattering properties and porosity of ices have to be quantified along with their THz--IR response in order to adequately interpret astrophysical observations. The developed model paves the way for solving this demanding problem of laboratory astrophysics.}

\maketitle

\section{Introduction}
\label{SEC:Intro}

Solving many problems in astrophysics and astrochemistry 
requires knowledge of the physical properties of interstellar and circumstellar ices \citep{ARAA.52.1.541.2015}.
These problems include, in particular, the evolution of molecular clouds and the formation of stellar systems \citep{Dominik_Tielens_1997ApJ, Oberg_Bergin_2021, Caselli_ApJ_2022}, and the formation of new molecular species in space and their abundances \citep{FASS.8.757619.2021, ARAA.58.1.727.2020, Mifsud2021}.
Measurements of the continuum emission in the millimeter, terahertz (THz), and far-IR ranges \citep{ARAA.58.1.727.2020, ARAA.57.1.79.2019} are major sources of observational data for astrophysical ices. 
Hence, there is a need for equivalent laboratory measurements and the resulting databases of broadband optical properties for different analogs of interstellar and circumstellar ices, including 
water (\ce{H2O}),
carbon monoxide (\ce{CO}),
and carbon dioxide (\ce{CO2}),
as well as a few other main constituents and their mixtures \citep{ARAA.57.1.79.2019, AA.629.A112.2019, 2022A&A...667A..49G, PCCP.16.8.3442.2014, FD.168.461.2014, PCCP.18.30.20199.2016}.

Most of the ongoing laboratory research is concentrated on studying astrophysical ice analogs of different compositions, while substantially fewer efforts have been put into obtaining information on structural features, such as porosity. 
These features depend on the experimental methodology of growing ice films, including different approaches and conditions (i.e., temperature, pressure, growth rate, etc.) of depositing ice on a cold substrate from the gas phase \citep{Cartwright2008, Loeffler2016, Millan2019}.
Annealing at different temperatures affects the structure of ice samples, promoting phase transition \citep{Isokoski2014, Cazaux2015, Schiltz2024}.
In the laboratory, the thickness of ice samples is usually varied between $\sim 100$~nm and $10$~$\mu$m for IR measurements, and from $\sim 10$~$\mu$m to $1$~mm for THz and millimeter measurements, with the aim of ensuring the spectroscopic measurements are sufficiently sensitive
\citep{2022A&A...667A..49G}.
Ice porosity generally increases with the thickness, ranging from several to tens of percent by volume \citep{Millan2019, Loeffler2016, Westley1998}.
High porosity impacts the THz--IR response of laboratory ices, leading to a reduction in the effective optical constants \citep{Millan2019, Loeffler2016}, a broadening of absorption peaks, and a reduction in their amplitude \citep{Loeffler2016, Schiltz2024}, or even to the dielectric spectrum redistribution and formation of additional absorption bands due to crystalline lattice disordering, such as the Boson peaks \citep{2022A&A...667A..49G, PRL.91.20.207601.2003, RMP.72.3.873.2000}.
At higher frequencies, porosity induces a strong scattering of electromagnetic waves, which makes ice samples optically opaque. This effect restrains the spectral range of ice characterization, leading to profound differences between the THz--IR responses of compact and porous samples, which should be taken into account in  both laboratory measurements and subsequent calculations of radiative transfer in the interstellar medium \citep[ISM;][]{Mitchell2017, Rocha2024, Schiltz2024}.

While it is broadly believed that porous ices are rare in the ISM \citep{Keane2001} because of their thermal annealing, exposure to cosmic rays and vacuum ultraviolet, or H-atom bombardment \citep{Palumbo2006, Raut2007, Raut2008, Palumbo2010, Accolla2011}, there is still no complete consensus on this issue.
Relying on the available observations, it is challenging to determine the actual porosity of interstellar and circumstellar ices.
Among the main arguments in favor of compact ices is the common absence of dangling OH features, which trace unbound water molecules. However, some laboratory data have also revealed the absence of such modes for porous samples \citep{Raut2007, Isokoski2014, He2022}.
At the same time, the recent work by \cite{Noble2024} shows a detection of the dangling OH features with the JWST NIRCam in Chamaeleon~I, which could be an indication of a porous ice structure. 

If porous ices were 
present in the ISM, their structural properties would play an important role in a number of processes, such as accretion, desorption, segregation, and diffusion. 
Because of the larger effective surface area, porous ices are expected to be more chemically reactive, enabling the trapping of volatiles and the freezing out of additional atoms and molecules.
The possible existence of porous ices in the ISM would stimulate the formation of various complex organic molecules, photodesorption processes \citep{Oberg2009_AA_1, Oberg2009_AA_3, Fayolle2011}, and exothermic solid-state reactions \citep{Cazaux2010, Dulieu2013}.
Thereby, knowledge of the structural features and porosity of ices from laboratory experiments, as well as from the related models of their dielectric response, is needed to assist in the interpretation of astronomical observations \citep{Westley1998, Accolla2011, Bossa2014, Isokoski2014, Loeffler2016, Millan2019, Rocha2024}.

A few available works on the porosity of astrophysical ice analogs present results for \ce{H2O}, \ce{CO}, \ce{CO2}, and their mixtures. Often, a quartz-crystal microbalance is used to estimate the density of ices, while interferometry is applied to quantify their refractive index in the visible and near-IR ranges \citep{Westley1998, Loeffler2016, Millan2019}. 
For this purpose, information about the density and refractive index of bulk is usually taken from the literature 
\citep{1926RSPSA.111..224M, Schulze1980, Warren1986, Westley1998, Dohnalek2003}. 
Another option is to use diffraction measurements to directly obtain the density of bulk regions in a porous sample 
\citep{Millan2019}.
Although the effective medium theory (EMT) makes it possible to estimate porosity from spectroscopic data \citep{Bossa2014, Millan2019}, it requires a priori knowledge of the properties of compact ices, typically taken from the literature.
Additional IR measurements can be applied to check the specific modes expected for porous ices \citep{Bossa2014, Isokoski2014, Mitchell2017, Nagasawa2021, Rocha2024}.
Less frequently, scanning electron microscopy \citep{Cartwright2008}, Monte Carlo simulations \citep{Cazaux2015, Clements2018}, and other experimental and theoretical techniques are applied to study the morphology of laboratory ices.

This series of papers is focused on measurements and analysis of the broadband optical properties of laboratory ices.
In \citet{AA.629.A112.2019}, we developed the experimental setup and methods to grow laboratory ice samples, measure their THz complex transmission (using the THz pulsed spectrometer), and retrieve their THz optical properties from the measured data.
In \citet{2022A&A...667A..49G}, we describe the method for merging the data of THz pulsed spectroscopy and Fourier-transform IR (FTIR) spectroscopy for the broadband characterization of ices, which was then applied to study the THz--IR response of \ce{CO} and \ce{CO2} ices.
In these broadband measurements, the high-frequency edge of the analyzed spectral bands was limited to $12$~THz \citep{2022A&A...667A..49G}; higher frequencies were excluded from the analysis due to monotonically increasing extinction, attributed to scattering.
Also, for both \ce{CO} and \ce{CO2} ices, we observe low-intensity  absorption peaks at higher frequencies that originate from a disordered crystal lattice and disappear upon annealing \citep{2022A&A...667A..49G}.
These spectroscopic features further motivated our research into the effects of scattering and porosity.

In the present paper, we develop a model to derive the relationship between the THz--IR optical properties of ices and their porosity.
A combined use of the Bruggeman EMT, Lorentz-Mie, and Rayleigh scattering theories, as well as the radiative transfer theory, enabled the characterization of both the reduced effective optical properties of porous ices (occurring at all frequencies) and the additional wave extinction due to scattering on pores (which increases with frequency). 
This model shows that scattering in the studied ice samples occurs mainly in the Rayleigh regime for frequencies of up to $32$~THz.
We conclude that pores of different shapes and dimensions can be approximated by spheres of effective radius.
The scattering properties and porosity were determined by comparing the measured broadband response of our \ce{CO} and \ce{CO2} laboratory ices with the literature data on reportedly compact ices. 
The porosity of the \ce{CO} and \ce{CO2} ice samples in our experiments is about $15$\% and $22$\%, respectively. Neglecting the porosity in the data analysis leads to a proportional underestimate of the optical properties. 

\section{Methods}
\label{SEC:Methods}

\subsection{THz--IR data on
porous \ce{CO} and \ce{CO2} ices}

To analyze scattering characteristics and porosity of \ce{CO} and \ce{CO2} ices, we used their THz--IR optical properties as reported by \cite{2022A&A...667A..49G} for a limited spectral range below $12$~THz.
In the present study, we extended the spectral range to $32$~THz, which allowed us to take scattering fingerprints into account.
Details on the ice growth process and THz--IR measurements are presented in \cite{AA.629.A112.2019} and \cite{2022A&A...667A..49G}.

Given the apparent scattering signatures seen in the collected data, we concluded that the studied ices are porous and thus kept in mind the following two main effects peculiar to porous analytes:
\begin{itemize}
\item In the entire spectral range, porous samples possess reduced refractive index and absorption coefficient (compared to that of a compact sample). These are referred to as the effective optical properties commonly defined within the EMT theory \citep{AO.20.1.26.1981, PQE.62.1.2018, Millan2019}.
\item 
The scattering increases monotonically with frequency until the analyte becomes opaque. The total extinction is usually defined as a superposition of the power absorption and scattering coefficients, $\mu_\mathrm{total} = \mu_\mathrm{\alpha} + \mu_\mathrm{s}$ \citep{Bohren1998}.
\end{itemize}

\noindent Below we treat the THz--IR results reported earlier for \ce{CO} and \ce{CO2} ices as effective optical constants of the respective compact ices, and summarize the methods to describe the aforementioned effects and to quantify the scattering properties,
porosity, and effective pore radius of ices.

\subsection{The Bruggeman EMT model}
\label{SEC:EMT}

Following our previous papers
\citep{AA.629.A112.2019, 2022A&A...667A..49G},
we describe the THz--IR response of ices
by the refractive index ($n$)
and absorption coefficient $\alpha$ (by field). These are related to
the complex refractive index
$\tilde{n} = n' - i n''$
and complex dielectric permittivity
$\tilde{\varepsilon}
= \varepsilon' - i \varepsilon''$ via $\tilde{n}
= n - i \left( c_\mathrm{0} / 2 \pi \nu \right) \alpha
= \sqrt{ \tilde{\varepsilon} }$,
where $n \equiv n'$, $\nu$ is the frequency,
and $c_\mathrm{0}$ is the speed of light in vacuum.
We point out that
the power absorption coefficient
(within the radiative transfer theory)
is given by $\mu_\mathrm{\alpha} = 2 \alpha$. 

\begin{figure}[!b]
    \centering
    \includegraphics[width=1.0\columnwidth]{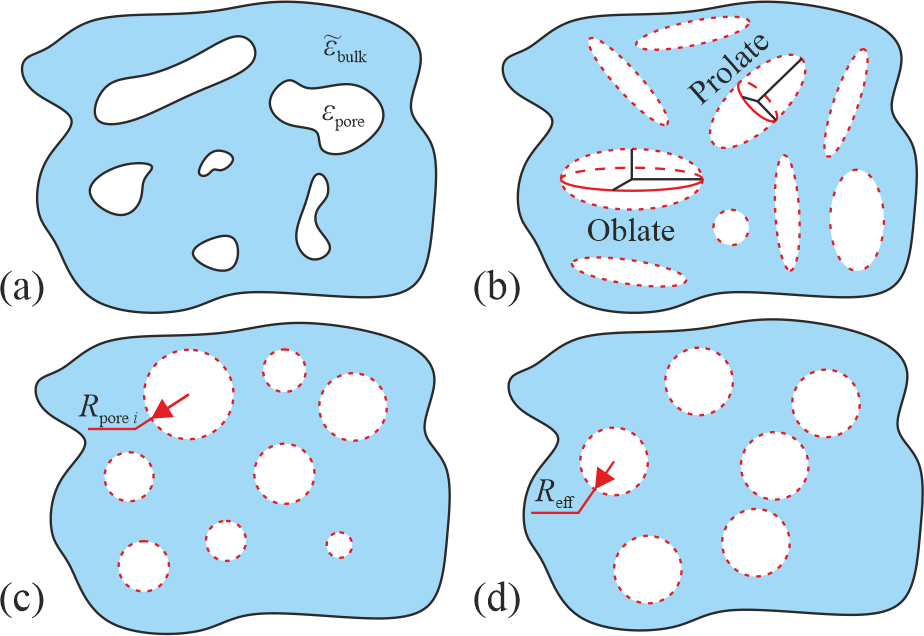}
    \caption{Pores of arbitrary shapes  (a) can be approximated by spheroids  (b) or spheres (c) of different sizes, or spheres of the same effective radius, $R_\mathrm{eff}$ (d).}
    \label{FIG:IceStructure}
\end{figure}

To define a relation
between the compact and porous analytes
at a given THz--IR frequency,
we used the Bruggeman EMT model
\citep{Bruggeman1935},
derived for the optically isotropic two-component mixtures of
compact host medium
and sub-wavelength ($\ll \lambda$) spheres.
Although this model is considered to be applicable
for spherical particles 
with the volume fraction of
$ 1/3\leq P\leq 2/3$
\citep{AP.138.1.78.1982, PQE.62.1.2018},
it appears to be relevant also for
particles of a more complex shape
and a wider range of volume fractions
(up to $ 0\leq P\leq 1$).
In particular, the model was shown to be
suitable for describing
the dielectric response of complex biological objects
\citep{Optica.8.11.1471.2021}
and porous materials
\citep{AO.59.28.8822.2020, OME.10.9.2100.2020, OME.12.8.3015.2022}.
Also, it can be extended to describe
the response of multicomponent systems
\citep{PQE.62.1.2018}.
If an analyzed medium is a compact material filled with empty pores,
Fig.~\ref{FIG:IceStructure}a,
the model is given by
\begin{equation}
    \left( 1 - P \right) \frac{ \tilde{\varepsilon}_\mathrm{bulk} -\tilde{\varepsilon} }{\tilde{\varepsilon}_\mathrm{bulk} + 2 \tilde{\varepsilon} } + P \frac{ \varepsilon_\mathrm{pore} - \tilde{\varepsilon} }{\varepsilon_\mathrm{pore} + 2 \tilde{\varepsilon} }
    = 0,
    \label{EQ:Bruggeman}    
\end{equation}
where $\tilde{\varepsilon}$
is the effective complex dielectric permittivity of a porous medium, 
$\varepsilon_\mathrm{pore} = 1$ is the dielectric permittivity of pores (empty space)
and $P$ is their volume fraction, while
$\tilde{\varepsilon}_\mathrm{bulk}$ is the dielectric permittivity of the compact material.

In fact, considering some prior knowledge on
the structural properties of an analyte,
one can also apply other EMT models of
complex dielectric permittivity
\citep{PU.50.6.595.2007, PQE.62.1.2018},
such as the Maxwell-Garnett
\citep{JOSAA.33.7.1244.2016}
and Landau-Lifshitz-Looyenga
\citep{LL.Book.1984, Phys.31.3.401.1965}
approximations,
equation for porous composites proposed by
\cite{AP.138.1.78.1982},
and simple linear decomposition of dielectric spectra
\citep{PMB.61.18.6808.2016}.
At a level set by
typical $\sim 1$\% confidence interval for
spectroscopic measurements,
predictions of these EMT models are very close
for a wide range of
dimensions, dielectric constants, and volume fractions of scatterers,
with the Bruggeman and Landau–Lifshitz–Looyenga models
providing somewhat better fits
\citep{Millan2019, AO.59.13.D6.2020}.
Dealing with optically anisotropic media,
one can apply the approximation of elliptical particles
\citep{Phys.12.5.257.1946}
or Lichtenecker
model \citep{PZ.27.115.1926, PZ.32.255.1931, IEEETMTT.58.3.545.2010},
since structural anisotropy and the resulting birefringence
can considerably affect the measured data
\citep{APR.2.1.2000024.2021, APLPhot.7.7.071101.2022, SR.13.16596.2023}.

\subsection{Modeling THz--IR scattering in ices}

Laboratory ices deposited from gas phase on a cold substrate are intrinsically porous. The structure of pores is generally quite complex and their dimensions can vary broadly, as sketched in Fig.~\ref{FIG:IceStructure}a.
A prior knowledge or assumptions
about characteristic size and volume fraction of pores
as well as about the dielectric contrast between the pores and compact ice
are required to adequately select
the level of simplification
and then accurately model
the radiative transfer.
Geometry of pores is commonly simplified
to spheroids or spheres of different dimensions; see
Figs.~\ref{FIG:IceStructure}b and~\ref{FIG:IceStructure}c, respectively.

Although our knowledge on
the porosity of ices is limited,
some robust assumptions can be introduced.
\begin{itemize}
\item First, laboratory ices are usually transparent
in a wide spectral range
spanning the THz and IR bands,
down to the wavelengths of $\lambda \sim 1$--$10$~$\mu$m
\citep{2022A&A...667A..49G}.
This implies fairly small pores compared to
the wavelength scale
($R_\mathrm{pore} \ll \lambda$). 
In such cases, the Rayleigh theory
can be applied to accurately describe the scattering properties of ices
\citep{Bohren1998}.
\item Second, the average pore size cannot be smaller
than the measured diffusion length in the ice.
For diffusion of various gases in amorphous solid water ice,
this length can be as small as $\sim 10$~ nm
\citep{Furuya2022},
while similar values are expected
for ices of other molecular compositions.
This exceeds typical scales of ice molecules
or crystalline lattices,
suggesting the minimum size scale for a single pore.
\end{itemize}

We assumed the intrinsic absorption of compact ice to be negligible 
in the spectral range of the scattering analysis
\citep[otherwise, the scattering coefficient cannot by rigorously defined; see][]{JQSRT.206.241.2018, PRA.20.5.054050.2023}.
Also, we used the single scattering approximation
that is relevant in the case of small volume fractions of pores
and low dielectric contrast between the scatterer and the host medium.
Then one can write
the scattering coefficient $\mu_\mathrm{s}$ for a given wavelength
in the following form:
\begin{equation}
    \mu_\mathrm{s} = \int_{V_\mathrm{min}}^{V_\mathrm{max}} 
    \frac{d \rho \left( V \right) }{ d V } C_\mathrm{sca} \left( V \right) dV,
    \label{EQ:muRayInt}    
\end{equation}
where $d \rho \left( V \right) / d V$
is the unknown volume distribution of pores,
$C_\mathrm{sca} \left(V \right) = c_{\rm sca} V^2$
is the Rayleigh scattering cross section with $c_{\rm sca}$ describing the wavelength and shape dependence (see Sect.~\ref{SEC:Rayleigh}),
$V_\mathrm{min}$ and $V_\mathrm{max}$
are the minimum and maximum volumes of pores. Assuming the volume distribution has a power-law spectrum with the exponent $-\gamma$, it can be conveniently normalized 
to the porosity ($P$; which is the first moment of the distribution),
\begin{equation}
    \frac{ d \rho \left( V \right) }{ d V } \approx P \frac{ 2 - \gamma }{ V_\mathrm{max}^2 }
    \left(\frac{ V }{ V_\mathrm{max} } \right)^{-\gamma},
    \label{EQ:DiffDensity}    
\end{equation}
where condition $\gamma < 2$ ensures that the value of $P$ is dominated by bigger pores.
Then the scattering coefficient takes the form
\begin{equation}
    \mu_\mathrm{s} \approx \frac{ 2 - \gamma}{ 3 - \gamma} P  V_\mathrm{max} c_\mathrm{sca} 
     \equiv \frac{ 2 - \gamma}{ 3 - \gamma} N_\mathrm{V} C_\mathrm{sca} \left( V_\mathrm{max} \right),
    \label{EQ:muRayDistibution}    
\end{equation}
where $C_\mathrm{sca}$ is calculated for a pore of maximum volume $V_\mathrm{max}$
and $N_\mathrm{V} \equiv P / V_\mathrm{max}$
is the number of such pores per unit volume.
Hence, the scattering effects are dominated by larger pores.

We see that, even without a prior knowledge about ice porosity,
we can limit our analysis to effective characterization of large-scale pores
of the volume $V_\mathrm{eff}\propto R_\mathrm{eff}^3$; see Fig.~\ref{FIG:IceStructure}d.
For such pores,
the scattering coefficient is reduced to the common form
\citep{Bohren1998}:
\begin{equation}
    \mu_\mathrm{s} = \left. N_\mathrm{V} C_\mathrm{sca}\right|_{V_\mathrm{eff}}\propto V_{\rm eff},
    \label{EQ:muRaySimpl}    
\end{equation}
where the unknown factor $\left( 2 - \gamma \right) / \left( 3 - \gamma \right)\sim1$ present in Eq.~\eqref{EQ:muRayDistibution} is absorbed in $V_\mathrm{eff}$.

Next, we summarize essential facts regarding
the Rayleigh and Lorentz-Mie scattering theories
and the related approaches to defining scattering parameters
that are required to analyze the ice porosity.

\subsubsection{Scattering cross section
and scattering coefficient
based on the Rayleigh approximation}
\label{SEC:Rayleigh}

In the Rayleigh limit,
the scattering cross section $C_\mathrm{sca}$
in Eq.~\eqref{EQ:muRaySimpl}
depends on the pore polarizability $\alpha_\mathrm{p}$ as
\begin{equation}
    C_\mathrm{sca} = \frac{ k^4 }{ 6 \pi } \left| \alpha_\mathrm{p} \right|^2,
    \label{EQ:Csca}    
\end{equation}
where $k = 2 \pi n_\mathrm{bulk} / \lambda $
is the wavenumber in a bulk medium.
For spherical pores of
radius $R_\mathrm{pore}$ (see
Fig.~\ref{FIG:IceStructure}c),
the polarizability $\alpha_\mathrm{p}$ is
\begin{equation}
    \alpha_\mathrm{p}
    = 4 \pi  R_\mathrm{pore}^3 \frac{ \varepsilon_\mathrm{pore} - \tilde{\varepsilon}_\mathrm{bulk}}{\varepsilon_\mathrm{pore} + 2 \tilde{\varepsilon}_\mathrm{bulk}},
\label{EQ:alphaSp}    
\end{equation}
while for spheroidal pores with semi-axes $a$, $b$, $c$ and the equivalent radius $(abc)^{1/3}$ (see Fig.~\ref{FIG:IceStructure}b), it is defined by the polarizability components along the principal axes,
\begin{equation}
    \alpha_{\mathrm{p}\,i} = 4 \pi a b c \frac{ \varepsilon_\mathrm{pore} - \tilde{\varepsilon}_\mathrm{bulk} }{ 3 \tilde{\varepsilon}_\mathrm{bulk} + 3 L_i \left( \varepsilon_\mathrm{pore} - \tilde{\varepsilon}_\mathrm{bulk} \right)},
    \label{EQ:alphaEll}
\end{equation}
where $L_i$ are the corresponding geometrical factors and $\sum_{i=1}^{3} L_i = 1$. For prolate pores with the semi-axes $a>b=c$ and the geometrical factors $L_2 = L_3$, we have
\begin{equation}
    \begin{split}
        L_1 &= \frac{ 1 - e^2 }{ e^2 } \left[ - 1 + \frac{ 1 }{ 2 e } \ln \left( \frac{ 1 + e }{ 1 - e } \right) \right],\\
        e^2 &= 1 - \frac{ c^2 }{ a^2},
    \end{split}
    \label{EQ:Lprolate}    
\end{equation}
while for oblate pores with $a = b > c$ and $L_1 = L_2$,
\begin{equation}
    \begin{split}
        L_3 &= \frac{ 1 + e^2 }{ e^3 } \left( e - \arctan e \right),\\
        e^2 &= \frac{ a^2 }{ c^2 } - 1.
    \end{split}
    \label{EQ:Loblate}    
\end{equation}
For randomly oriented spheroids,
the average scattering cross section
is given by
\begin{equation}
    \langle C_\mathrm{sca} \rangle
    = \frac{ k^4 }{ 6 \pi } \left( \frac{ 1 }{ 3 } \sum_{i=1}^{3} \left| \alpha_{\mathrm{p}\,i} \right|^2 \right).
    \label{EQ:CscaEll} 
\end{equation}

In the Rayleigh approximation, 
the scattering coefficient
is proportional
to the fourth power of frequency,
$\mu_\mathrm{s} = A \nu^4$,
where $A$ is a function of
the relative dielectric permittivity of compact ice $\tilde{\varepsilon}_\mathrm{bulk}/\varepsilon_{\rm pore}$, 
the pore shape and 
size distribution,
and the porosity ($P$).
In this way,
analysis of experimental spectroscopic data
makes it possible to estimate $A$
and thus to obtain information about porosity.

\subsubsection{Scattering cross section and scattering coefficient based on rigorous Lorentz-Mie theory}

The cross section of a spherical pore of arbitrary radius $R_\mathrm{pore}$ (see Fig.~\ref{FIG:IceStructure}c) is defined within the general Lorentz-Mie scattering theory \citep{Bohren1998},
\begin{equation}
C_\mathrm{sca}
    = \frac{2\pi}{k^2} \sum_{n=1}^{\infty} \left( 2n+1 \right) \left( \left| a_n \right|^2 + \left| b_n \right|^2 \right),
\label{EQ:CscaMie} 
\end{equation}
where
\begin{equation}
    \begin{split}
    a_\mathrm{n} &= \frac{ m \psi_\mathrm{n} \left( m x \right) \psi'_\mathrm{n} \left( x \right) - \psi_\mathrm{n} \left( x \right) \psi'_\mathrm{n} \left( m x \right) } { m \psi_\mathrm{n} \left( m x \right) \xi'_\mathrm{n} \left( x \right) - \xi_\mathrm{n} \left( x \right) \psi'_\mathrm{n} \left( m x \right) },\\
    b_\mathrm{n} &= \frac{ \psi_\mathrm{n} \left( m x \right) \psi'_\mathrm{n} \left( x \right) - m \psi_\mathrm{n} \left( x \right) \psi'_\mathrm{n} \left( m x \right) }{ \psi_\mathrm{n} \left( m x \right) \xi'_\mathrm{n} \left( x \right) - m \xi_\mathrm{n} \left( x \right) \psi'_\mathrm{n} \left( m x \right)}
    \end{split}
    \label{EQ:CoeffMie}
\end{equation}
are the Mie scattering coefficients,
$\psi_\mathrm{n} \left( x \right)$ and
$\xi_\mathrm{n} \left( x \right)$
are the Riccati-Bessel functions,
$m = \varepsilon_\mathrm{pore} / \tilde{\varepsilon}_\mathrm{bulk}$,
and $x = k R_\mathrm{pore}$.
Relying on the Lorentz-Mie cross section,
the scattering coefficient $\mu_\mathrm{s}$
is usually defined 
using Eq.~\eqref{EQ:muRaySimpl}.
We should stress, however, that such a definition
should be used cautiously 
-- in fact, it is 
only applicable if scattering pores are still quite small
and their scattering phase functions
are overall isotropic 
\citep[i.e., the Rayleigh-like; see][]{PRA.20.5.054050.2023}.
Otherwise,
the parameter $\mu_\mathrm{s}$
loses its physical meaning
and is no longer relevant to describe
the radiative transport in porous media.

The limits of applicability of
the aforementioned Rayleigh
and Lorentz-Mie scattering theories
(in terms of
the pore sizes
and the character of scattering phase function)
depend on a number of factors,
and are commonly examined experimentally
or by numerical analysis
\citep{PRA.20.5.054050.2023}.

\section{Results}
\label{SEC:Results}

The scattering effects
were observed in our previous measurements of \ce{CO} and \ce{CO2} ices
at frequencies above
$\nu = 12$~THz,
corresponding to the wavelengths below
$\lambda \approx 25$~$\mu$m
\citep{2022A&A...667A..49G}.
These ice films were transparent or translucent
at frequencies up to $\nu = 32$~THz
(or at wavelengths down to $\lambda \approx 9.4~\mu$m).
Above the frequency of $18$~THz,
some narrow frequency bands are excluded from the analysis
due to the non-transparency of a beamsplitter
in our FTIR spectrometer.
Even though quantification of a broadband complex dielectric permittivity
based on IR data and the Kramers-Kronig transform
\citep{2022A&A...667A..49G}
is generally problematic if some spectral fragments are missing,
in the present study we obtained robust estimates for
the ice scattering coefficient
at frequencies up to $32$~THz
(while neglecting some spectral bands
with no FTIR signal and intense absorption by ice).
Such scattering can be induced by pores
of sizes much smaller than
the aforementioned wavelengths, i.e., $\lesssim 1$~$\mu$m,
which highlights the need to justify applicability of the Rayleigh scattering regime, as elaborated below.

\subsection{Approximating pores using effective spheres}

Since the actual shape of pores is unknown,
we calculated
the Rayleigh scattering cross sections
$C_\mathrm{sca}$
for a sphere of volume $\frac{4}{3} \pi R_\mathrm{pore}^3$ and for spheroids with varying $c/a$ ratio but fixed volume 
$\frac{4}{3} \pi abc$, equal to that of a sphere.
In our analysis,
the dielectric permittivity of pores
corresponds to that of free space
($\varepsilon_\mathrm{pore} = 1$),
while that of bulk host medium is set to
$\varepsilon_\mathrm{bulk} = 2$,
which is reportedly close to the value for compact \ce{CO2} ice
\citep{Warren1986, Gerakines2020}.

\begin{figure}[!t]
    \centering
    \includegraphics[width=1.0\columnwidth]{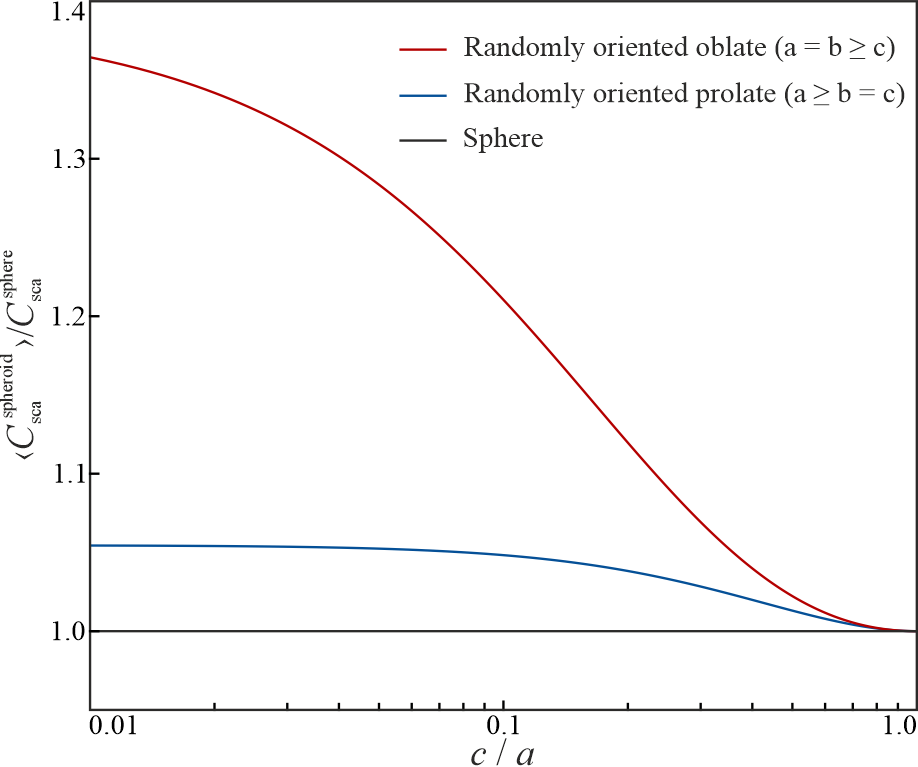}
    \caption{Ratio of
    the Rayleigh scattering cross section
    of a randomly oriented oblate or prolate spheroid,
    $\langle C_\mathrm{sca}^\mathrm{spheroid} \rangle$,
    to that of a sphere 
    of the same volume, $C_\mathrm{sca}^\mathrm{sphere}$.
    The dielectric permittivity
    of pores and host medium
    are $\varepsilon_\mathrm{pore} = 1$
    and $\varepsilon_\mathrm{bulk} = 2$, respectively.}
    \label{FIG:SphEll}
\end{figure}

Figure~\ref{FIG:SphEll}
shows the scattering cross section 
$\langle C_\mathrm{sca}^\mathrm{spheroid} \rangle$
of randomly oriented prolate and oblate spheroids, normalized to that of a sphere $C_\mathrm{sca}^\mathrm{sphere}$ and plotted versus the ratio $c/a$.
We notice that the difference
between the scattering cross section
of prolate spheroids and a sphere
is less than $5\%$
in a wide range of $c/a = 0.01$--$1.0$,
while that of oblate spheroids and a sphere
is less than $10\%$ for $c/a = 0.25$--$1.0$.
Thus, when such anisotropic pores are randomly oriented, their net contribution is essentially determined by their volume, and the 
scattering cross section $C_\mathrm{sca}$
can be reasonably approximated by that of
a sphere.

Regarding our laboratory ice films
\citep{2022A&A...667A..49G},
they were assumed to be isotropic
due to the diffuse regime of ice growth,
attributed to a sufficiently high deposition pressure in the chamber
\citep{AA.629.A112.2019}.
This allowed us to exclude
any residual anisotropy of pores
from our further analysis
and treat them as equivalent spheres.

\subsection{Justifying the Rayleigh scattering regime}

\begin{figure}[!t]
    \centering
    \includegraphics[width=1.0\columnwidth]{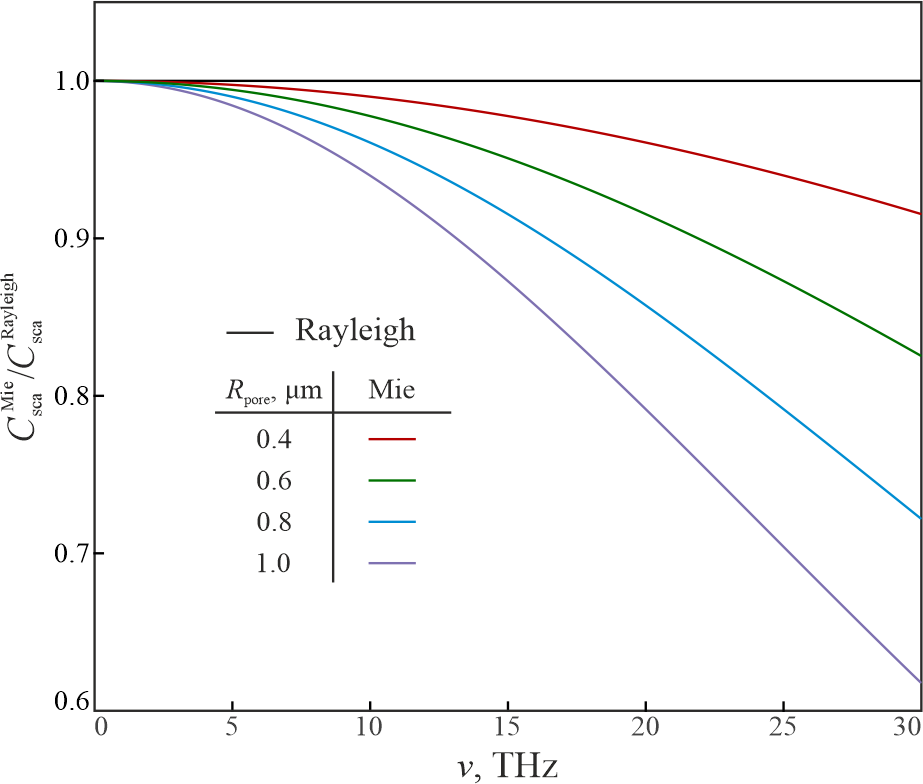}
    \caption{Frequency-dependent ratio
    of the Lorentz-Mie ($C_\mathrm{sca}^\mathrm{Mie}$)
    and Rayleigh ($C_\mathrm{sca}^\mathrm{Rayleigh}$)
    scattering cross sections for spherical pores
    with $\varepsilon_\mathrm{pore} = 1$,
    embedded in a host medium
    with $\varepsilon_\mathrm{bulk} = 2$. Different curves show the results for $R_\mathrm{pore} = 0.4$, $0.6$, $0.8$, and $1$~$\mu$m.}
    \label{FIG:MieRay}
\end{figure}

Next, we compared the scattering cross section ($C_\mathrm{sca}$)
of spherical pores of radius $R_\mathrm{pore}$
obtained from the Rayleigh
[Eqs.~\eqref{EQ:Csca} and~\eqref{EQ:alphaSp}]
and Lorentz-Mie
[Eqs.~\eqref{EQ:CscaMie} and~\eqref{EQ:CoeffMie}]
scattering theories.
In Fig.~\ref{FIG:MieRay},
the ratio of
the Lorentz-Mie $C_\mathrm{sca}^\mathrm{Mie}$
and Rayleigh $C_\mathrm{sca}^\mathrm{Rayleigh}$
scattering cross sections
is plotted in the frequency range of $0$--$30$~THz 
for pores of 
$R_\mathrm{pore} = 0.4$, $0.6$, $0.8$, and $1$~$\mu$m.
The discrepancy seen between the two theories naturally increases
with $\nu$ and $R_\mathrm{pore}$, which allowed us to describe crossover between the Rayleigh and Mie scattering regimes in our spectral data.
For frequencies $\nu<15$~ THz, where
the discrepancy is less than $10$\% for
$R_\mathrm{pore} \leq 0.8$~$\mu$m, 
one can claim that
the two theories give similar predictions.

The general Lorentz-Mie theory
is applicable for arbitrary ratios
of the pore radius $R_\mathrm{pore}$
and the wavelengths $\lambda$, while the Rayleigh approximation only
works in the limit of small pores,
$R_\mathrm{pore} \ll \lambda$.
Fortunately, values of the ice dielectric constants as well as ranges of 
frequencies and pore sizes relevant for our laboratory ices are such that
the Rayleigh theory can still be used.
In the following, we use it
to model the scattering cross section
$C_\mathrm{sca}$
(Eqs.~\eqref{EQ:Csca} and~\eqref{EQ:alphaSp}) and estimate the scattering coefficient $\mu_\mathrm{s}$
of laboratory ices
(Eq.~\eqref{EQ:muRaySimpl}).

\subsection{Porosity, scattering properties, and effective pore radius for laboratory ices}

\begin{figure*}[h!]
    \centering
    \includegraphics[width=1.9\columnwidth]{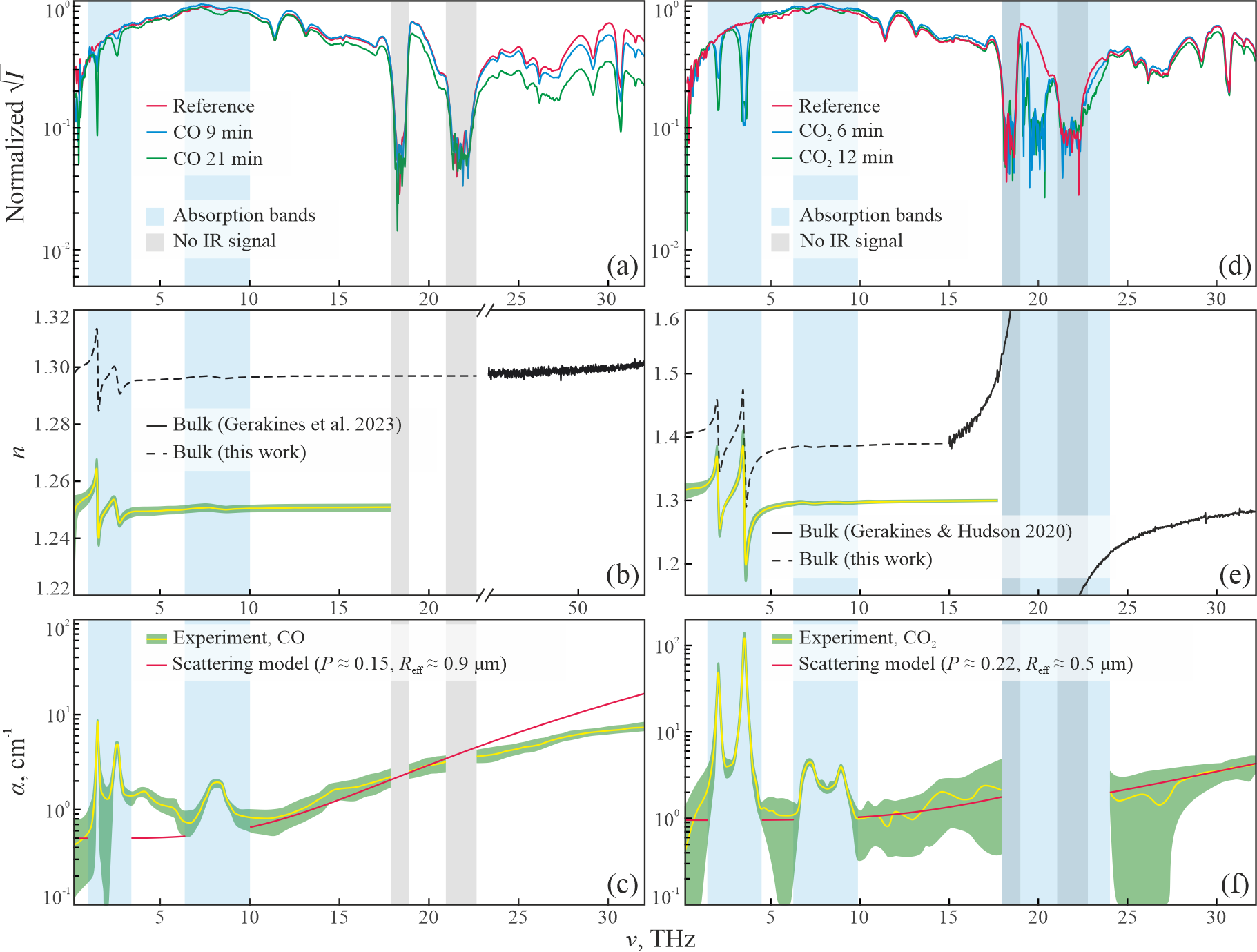}
    \caption{THz--IR optical properties of the studied \ce{CO} and \ce{CO2} ices along with the estimates of scattering parameters.
    Panel (a):~Examples of the reference and sample IR signals for \ce{CO} ices. The blue-shaded stripes indicate the absorption bands, and the gray-shaded stripes the ranges of low IR signal.
    Panels (b) and (c): Refractive index ($n$) and absorption coefficient ($\alpha$; by field, plotted in log scale) of our \ce{CO} ices.  We show the mean value (solid yellow lines) with the $\pm1.5\sigma$ confidence interval (green shading) for the measurements, as well as the calibrated value for $n$ (dashed black line). The results overlap with the literature data for high-frequency $n$ of compact \ce{CO} ice from \citet{Gerakines2023}, and with the modeled scattering coefficient $\mu_\mathrm{s}/2$, corresponding to $P \approx 15$\% and $R_\mathrm{eff} \approx 0.9$~$\mu$m.
    Panels (d)--(f):~Same but for \ce{CO2} ices. Here $\mu_\mathrm{s}/2$ corresponds to $P \approx 22$\% and $R_\mathrm{eff} \approx 0.5$~$\mu$m, and the literature data for compact \ce{CO2} ice are taken from \citet{Gerakines2020}.}
    \label{FIG:IcesScat}
\end{figure*}

The described methods of the EMT
and Rayleigh scattering theory
are applied to analyze
the scattering properties,
effective pore radii ($R_\mathrm{pore}$),
and porosity ($P$)
of our \ce{CO} and \ce{CO2} ices
\citep{2022A&A...667A..49G}. The results are summarized in Fig.~\ref{FIG:IcesScat}.
We excluded from the analysis
the aforementioned spectral bands of low signal
(gray-shaded areas),
as well as those with resonant absorption peaks
(blue-colored areas),
attributed to the vibrational modes of ices.

 In Fig.~\ref{FIG:IcesScat}b
we compare the THz--IR refractive index ($n$) of our \ce{CO} ices
at lower frequencies ($\leq 18$~THz)
with that at higher frequencies ($\gtrsim 45$~THz), reported in \cite{Gerakines2023} for thin compact samples.
Taking into account the fact that the
refractive index of dielectric materials shows increase
with decreasing frequency only 
while passing the spectral absorption features
\citep[as governed by the Kramers-Kronig relations; see][]{PR.161.1.143.1967},
and given the absence of pronounced absorption peaks for \ce{CO} ice
in the frequency gap between $18$ and $45$~THz,
we expect the refractive index of \ce{CO} ice
to be constant within this gap.
This allowed us to use the high-frequency refractive index of compact \ce{CO} ice from \citet{Gerakines2023}
to calibrate our THz--IR spectra.
In this way, the refractive indices
measured at $18$~THz and reported at $45$~THz
are used with the Bruggeman EMT model, Eq.~\eqref{EQ:Bruggeman},
to estimate the porosity of our \ce{CO} ice: the resulting value is found to be $P \approx 15$\%.
In Fig.~\ref{FIG:IcesScat}b
we also estimate the broadband refractive index of
compact $\ce{CO}$ ice
at lower frequencies $\lesssim 18$~THz
(using our experimental data for porous ice,
the retrieved value of porosity,
and the Bruggeman model),
as well as at higher frequencies
\citep[based on the data from][]{Gerakines2023}.

In Fig.~\ref{FIG:IcesScat}c we present the absorption coefficient $\alpha$ of our \ce{CO} ices. The curve increases monotonically with $\nu$ due to scattering.
It is fitted by one half of the scattering coefficient ($\mu_\mathrm{s}$) defined within the Rayleigh formalism,
\begin{equation}
    \mu_\mathrm{s} = A_\mathrm{1} + A_\mathrm{2} \nu^4,
    \label{EQ:mufit}    
\end{equation}
where $A_\mathrm{1}$ is a free parameter that accounts for noise level in our spectroscopic data \citep{2022A&A...667A..49G}, while $A_\mathrm{2}\propto PR_\mathrm{eff}^3$ is also a function of $\varepsilon_\mathrm{bulk}$.
The absorption coefficient $\alpha$ (by field)
is fitted by $\mu_\mathrm{s}/2$ in order to ensure consistency between our experimental data
and the Rayleigh theory.
One can see that experimental curve
is described well by the model at $\lesssim 23$~THz,
while at higher frequencies a growing deviation becomes evident.
The fitting procedure allowed us to quantify
the effective pore radius in $\ce{CO}$ ice,
which turned out to be as high as
$R_\mathrm{eff} \approx 0.9$~$\mu$m.
Such large pores cause cross-over
between the Rayleigh and Mie scattering regimes
at higher frequencies,
which explains the origin of the observed deviation.

\begin{figure*}[t!]
    \centering
    \includegraphics[width=2.0\columnwidth]{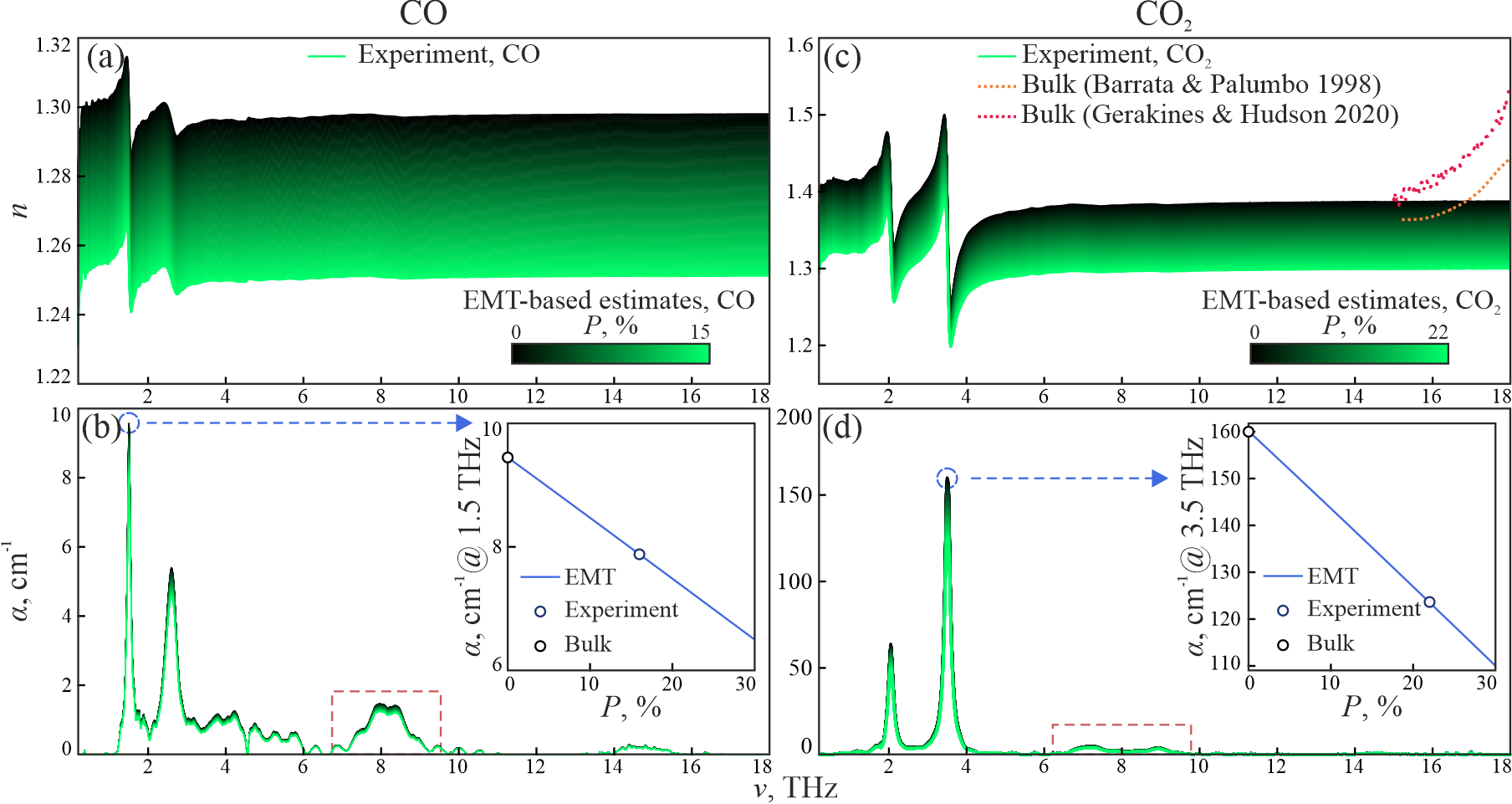}
    \caption{Bruggeman model-based predictions of
    the THz--IR optical properties of
    \ce{CO} and \ce{CO2} ices
    for different values of porosity ($P$).
    Panels (a) and (b): Refractive index ($n$)
    and absorption coefficient ($\alpha$; by field)
    of \ce{CO} ice. Panels
    (c) and (d): Same but for \ce{CO2} ice.
    The inserts in panels~(b) and~(d)
    depict the maximum amplitude of $\alpha$ for
    the most intense absorption peak
    in \ce{CO} (at $1.5$~THz)
    and \ce{CO2} (at $3.5$~THz), respectively,
    plotted versus $P$.
    The dashed red rectangles in panels~(b) and~(d)
    indicate the absorption peaks
    disappearing upon annealing (presumably caused 
    by the morphological features of porous ice).
    In panel~(c), the estimated refractive index of
    compact ($P=0$) \ce{CO2} ice 
    is compared with that of reportedly compact samples
    \citep{JOSAA.15.12.3076.1998, Gerakines2020}.}
    \label{FIG:EMT}
\end{figure*}

The results of similar analysis for $\ce{CO2}$ ices are presented in Figs.~\ref{FIG:IcesScat}e and~\ref{FIG:IcesScat}f.
Here, substantially more data on the THz--IR refractive index ($n$) of compact ices were reported earlier \citep{Warren1986, JOSAA.15.12.3076.1998, Gerakines2020}.
In panel~(e) we use the most recent results
\citep{Gerakines2020},
measured at higher frequencies
and overlapping with our spectra in the range of $15$--$18$~THz, which allowed us to calibrate our experimental data
and estimate the porosity of our $\ce{CO2}$ ice.
The derived value is $P \approx 22$\%,
which is even larger than that of our \ce{CO} ice.
In panel~(f) we fit the experimental absorption curve $\alpha$
with the theoretical scattering coefficient $\mu_\mathrm{s}/2$.
The resulting effective pore radius of
$R_\mathrm{eff} \approx 0.5$~$\mu$m
is, in turn, smaller than that derived for \ce{CO} ices.
The theory describes our experimental data for \ce{CO2} ices well,
justifying that scattering occurs in the Rayleigh regime
for frequencies below $32$~THz -- which, in turn, is in overall agreement with Fig.~\ref{FIG:MieRay}.

Summing up, the computed effective pore radii $R_\mathrm{eff}$ for
\ce{CO} and \ce{CO2} ices are in the submicron range,
and scattering occurs mostly in the Rayleigh regime.
Our estimates for the ice porosity ($P$)
are also quite reasonable, in a range from several to tens of percent, which is typical for laboratory ices
deposited from a gas phase on a cold substrate
\citep{Millan2019, Loeffler2016, Westley1998}.

\subsection{Optical properties of porous and compact laboratory ices}

The porosity of $P = 15$\% and $22\%$ derived for our \ce{CO} and \ce{CO2} ices, respectively, makes their responses considerably different from those of compact ices.
To quantify this difference for each ice, we modeled the response for $P$ in a range between the respective derived value and $P=0$ (compact ice), using
the Bruggeman model (Eq.~(\ref{EQ:Bruggeman})). Color-coded results for $n$ and $\alpha$ are plotted in Fig.~\ref{FIG:EMT}, showing 
an evident decrease in both $n$ and $\alpha$ with $P$. The inserts in panels~(b) and~(d) illustrate the modeled dependences for the most intense absorption peaks in \ce{CO} (at $1.5$~THz) and \ce{CO2} (at $3.5$~THz) ices, respectively. The Bruggeman model predicts an almost linear decrease in absorption, which is common for most EMT approximations \citep{PU.50.6.595.2007, PQE.62.1.2018, JOSAA.33.7.1244.2016, LL.Book.1984, Phys.31.3.401.1965, AP.138.1.78.1982, PMB.61.18.6808.2016}.

While the EMT has a potential to estimate the ice porosity and predict the optical properties of the bulk, as illustrated in Fig.~\ref{FIG:EMT}, it does not include several important effects underlying the response of porous samples with disordered crystalline lattice. 
In fact, the EMT is only able to predict the scaling dependences \citep{Millan2019, Loeffler2016}, while it does not account for redistribution of the absorption peaks \citep[such as changes in their magnitude, width, shape, and spectral position; see][]{Loeffler2016, Schiltz2024} and for the formation of new absorption contours \citep[inherent to disordered systems; see][]{2022A&A...667A..49G, PRL.91.20.207601.2003, RMP.72.3.873.2000}.
In Figs.~\ref{FIG:EMT}b and~\ref{FIG:EMT}d, such peaks can be seen near $\nu = 8.11$~THz for \ce{CO} ice, and near $\nu = 7.16$ and $8.88$~THz for \ce{CO2} ice: in \citet{2022A&A...667A..49G} they were shown to vanish upon low-temperature annealing, which leads to compaction of ices. We see that our model still recovers these peaks for $P=0$, which highlights limitations of the oversimplified EMT approaches.

In Fig.~\ref{FIG:EMT}c we compare the EMT-based predictions for the refractive index of compact \ce{CO2} ice with the high-frequency literature data reported for compact samples \citep{JOSAA.15.12.3076.1998, Gerakines2020}.
In contrast to the literature, our experimental curves and the EMT-based predictions do not show the dispersion at $15$--$18$~THz, which is caused by the side lobes of the known vibrational modes of \ce{CO2} ice near 20~THz.
This apparent discrepancy is simply a result of edge effects in the method applied to derive the optical constants \citep{2022A&A...667A..49G}: in the FTIR spectroscopy, the phase information (and, thus, the refractive index of the analyte) is retrieved from the transmission amplitude using the Kramers-Kronig transform. 
This requires the absorption features to be located well within the spectral range over which the integral is computed -- otherwise, the side effects can become dominant.

\section{Discussion and conclusion}

We have shown that unaccounted for ice porosity can lead to significantly underestimated optical constants in the THz--IR range. Considering the porosity values derived from our experiments as well as from other literature sources reporting on laboratory ices -- all in the range between a few to a few dozen percent \citep{Millan2019, Loeffler2016, Westley1998} -- we stress that such underestimations should be taken into account when using laboratory data to interpret astrophysical observations.

Our findings highlight the necessity for novel methods and protocols of laboratory ice growth and further thermal processing, which would facilitate ice compaction.
Our results also reveal the need for a better understanding the THz--IR response of ices in porous and compact states, including the development of adequate models of broadband complex dielectric permittivity of ices as a function of porosity.
Such models would be useful to describe continuum emission and radiative transfer in dense cold regions of the ISM, where thick icy mantles covering dust grains may 
be porous. Furthermore, porosity may 
increase with the mantle thickness, which is believed to vary over a broad range, from between a fraction of a nanometer to a fraction of a micron 
\citep{ARAA.52.1.541.2015},
and therefore the broadband optical properties of such mantles could vary dramatically too.

We point out that the method developed to analyze the ice porosity, based on the measured broadband optical properties, is quite generic. It can be useful in both laboratory astrophysics and other fundamental and applied branches of optical spectroscopy involving studies of porous samples.

To summarize, we have developed a model that yields a relationship between the THz--IR responses of compact and porous ices.
The model was used to analyze the measured THz--IR response
of \ce{CO} and \ce{CO2} laboratory ices, with the aim to estimate their scattering properties and porosity.
We have shown that the THz--IR scattering in laboratory ices occurs in the Rayleigh regime, and that the pores of different shapes and dimensions can be approximated by spheres of the same effective radius.
By comparing the measured responses of porous \ce{CO} and \ce{CO2} ices with the data available on compact ices, the porosity was found to be as high as $\approx15$\% and $\approx22$\%, respectively.
The porosity correction derived in the present paper should be used for the results presented in \citet{AA.629.A112.2019} and \citet{2022A&A...667A..49G}: neglecting it leads to an almost proportional relative error when estimating the THz--IR optical constants of compact ices.
Our findings suggest that the scattering properties and porosity of laboratory ices should be quantified and included in their THz--IR response, to correctly employ these data for interpreting astrophysical observations.

\begin{acknowledgements}
    A.V.I, B.M.G., and P.C. gratefully acknowledge the support of the Max Planck Society. The work of A.A.G., I.N.D., S.V.G., and K.I.Z. on processing of the THz pulsed spectroscopy  and FTIR data,
    analysis of porosity and scattering properties of ices
    was supported by the RSF project \#~$25$--$79$--$30006$.
\end{acknowledgements}

\end{document}